\documentstyle[12pt,    epsfig]{article}

\def\dl{D_L}
\def\daa{D_A}
\def\dpp{D_P}
\def\ov{\Omega_v}
\def\om{\Omega_m}
\def\b{\bigskip}
\def\Mo{M_{\oplus  }   }

\begin{document}

%%%%%%%%%%%%%%%%%%%%%%%%%%%%%%%%%%%%%%%%%%

\def\dl{D_L}
\def\daa{D_A}
\def\dpp{D_P}
\def\ov{\Omega_v}
\def\om{\Omega_m}
\def\b{\bigskip}
\def\Mo{M_{\oplus  }   } 
%%%%%%%%%%%%%%%%%%%%%%%%%%%%%%%%%%%%%%%%%%%%%%

\begin{center}
 { \bf 
OBSERVED BIMODALITY OF THE EINSTEIN CROSSING TIMES OF 
GALACTIC MICROLENSING
EVENTS
}
\end{center}

\b

\begin{center}
T Wickramasinghe\footnote[1]{Department of Physics, 
The College of New Jersey, Ewing, NJ 08628 }, 
M Struble\footnote[2]{Department of Physics \&\ Astronomy, 
University of Pennsylvania, Philadelphia, PA 19104}, and   J Nieusma$^1$  
\end{center} 

\b

\begin{center}
 { \bf 
ABSTRACT }
\end{center} 

\b

The OGLE data$^1$
 for Einstein ring crossing times, $t_E$, for microlensing
 events 
toward the galactic bulge  are analyzed. 
The analysis shows
 that the crossing times 
are bimodal, indicating that two populations of lenses 
could be  responsible for observed microlensing events. 
Given the possibility that microlensing in this 
direction can be due to both main-sequence 
stars and white dwarfs, we analyze and show that 
the observed bimodality  of $t_E$ can be derived from 
the accepted density distributions of both populations.
Our 
Kolmogorov-Smirnov (KS) one sample test 
  shows that 
 that a white dwarf population of 
about $25\% $ of all stars in the galaxy 
agrees well with the observed bimodality with a KS significance 
level greater than $97\%$. 

\section{Introduction}

The Einstein crossing time (measured in days)  of a galactic 
microlensing event can be written as$^2$
$$
t_E = 78.163  
\left(    
\frac{M  }{ M_{\odot }   }   
\right)^{\frac{1 }{2 }    } 
\left(    
\frac{ D_d}{ 10  \rm{kpc   }   }   
\right)^{\frac{1 }{2 }    } 
\left(    
1  -   \frac{ D_d } { D_s } 
\right)^{\frac{1 }{2 }    } 
\left(    
\frac{v    }{  200  \rm{ km  /s }  } 
\right)^{ - 1    } \, , 
\eqno(1) 
$$
where $M$ is the mass of the lens,  $v$ the relative 
 orbital 
speed of the lens and the source, and $D_d$ the distance of the lens, and 
$D_s$ the distance of the source. 

\b 

OGLE data for $190$ events (Fig. 1) show a certain bimodality in 
the Einstein crossing times. 
In order to reproduce this bimodality, we use Eq. (1) with 
respective galactic mass distributions of 
main-sequence stars and white dwarfs. 
Once we have chosen all the relevant parameters satisfying their 
respective density distributions, 
we carry out a Monte Carlo simulation 
to obtain a distribution for 
$t_E$. 
This distribution is then tested against the observed 
distribution utilizing the 
KS one sample test to investigate 
the agreement. 

\section{Analysis}

For our statistical analysis, the parameters in Eq. (1) were 
chosen as follows. 
About $90\%$ of $D_s$ values were chosen from the 
galactic bar. 
The density profile of the bar was assumed to be$^3$
$$
\nu(r_s) = \nu_0 \exp\left( -  \frac{1}{2} r_s^2 \right)  10^9  L_{ \odot } \; 
\rm{ pc}^{-3}  \, , 
\eqno(2)
$$
where 
$$
r_s  = \left\{
\left[
\left( \frac{ x^{\prime} }{x_0  }    \right)^2 
+
\left( \frac{ y^{\prime} }{y_0  }    \right)^2 
\right]^2
+
\left( \frac{ z^{\prime} }{z_0  }    \right)^4
\right\}^{ \frac{ 1 }{ 4 }   }  
\eqno(3)
$$
and $\nu_0 = 3.66  \times  10^7   L_{ \odot }$ kpc$^{-3}$.
The scale lengths are 
$x_0 = 1.85$ kpc, 
$y_0 = 0.62$ kpc, 
and
$z_0 = 0.43$ kpc, respectively. 
In these coordinates, the galactic center is at the origin.  
The rest of $D_s$ values were chosen from a disc population, 
whose density profile is given by 
$$
\rho_D = \rho_0  \exp  \left[  -  \frac{ \left|z^{\prime} \right|
}{h }    + 
\frac{ R_0 - s  }{ s_D } 
\right] \,   , 
\eqno(4) 
$$
where $\rho_0$ is the mass density in the local solar neighborhood, 
$h$ the scale height, and 
$s$ and $z^{\prime}$ form a system of 
cylindrical galactocentric coordinates.

\b

The lens distances $D_d$ were chosen only from the disk. 
The orbital speeds $v$ were chosen from a standard 
galactic rotation model.

\b

Both main-sequence stars and white dwarfs were used for masses. 
For main-sequences masses, a  combination of Scalo initial mass 
function$^4$ and a rapidly falling three-power law form 
due to Kroupa, et al.$^5$
was used as follows:
$$
\xi(M) \propto   \cases{
M^{- 2.35} & for  $ M  >   10 M_{\odot} $,   \cr 
M^{- 3.27} & for  $ 1 M_{\odot}   <   M  <  10   M_{\odot}   $,   \cr 
M^{- 2.2} & for  $ 0.5  M_{\odot}   <   M  <  1  M_{\odot}   $,   \cr 
M^{- 1.2} & for  $ 0.2   M_{\odot}   <   M  <  0.5     M_{\odot}   $,   \cr 
M^{- 1.85} & for  $ 0.1   M_{\odot}   <   M  <  0.2     M_{\odot}   $.   \cr 
}
$$

\b

Both DA and DB types of white dwarfs were used in our simulations. 
Their masses obey normal distributions such that 
$<M_{DA}> = 0.593  \pm 0.016 \,  M_{\odot} $ and 
$<M_{DB}> = 0.711 \pm 0.009  \,  M_{\odot} $ respectively.$^{6,7}$

\section{Analysis and Results}

Figure 1 shows the OGLE data and our simulated values of 
Einstein crossing times. 
The KS one sample test was used to see if the two distributions 
had been drawn from the same parent distribution. 
For any acceptable agreement, two different mass distributions 
were essential. 
In our calculations, these masses were chosen from 
both main-sequence stars and 
white dwarfs including both types DA and DB. 
We find that the white dwarf contribution should be as 
high as about $25\%$ to explain the data well. 
Out of this contribution, for a better agreement, we needed 
about $86\%$ of DA and about $14\%$ of DB dwarfs. 

\begin{figure}[h]
\centerline{\epsfig{file= 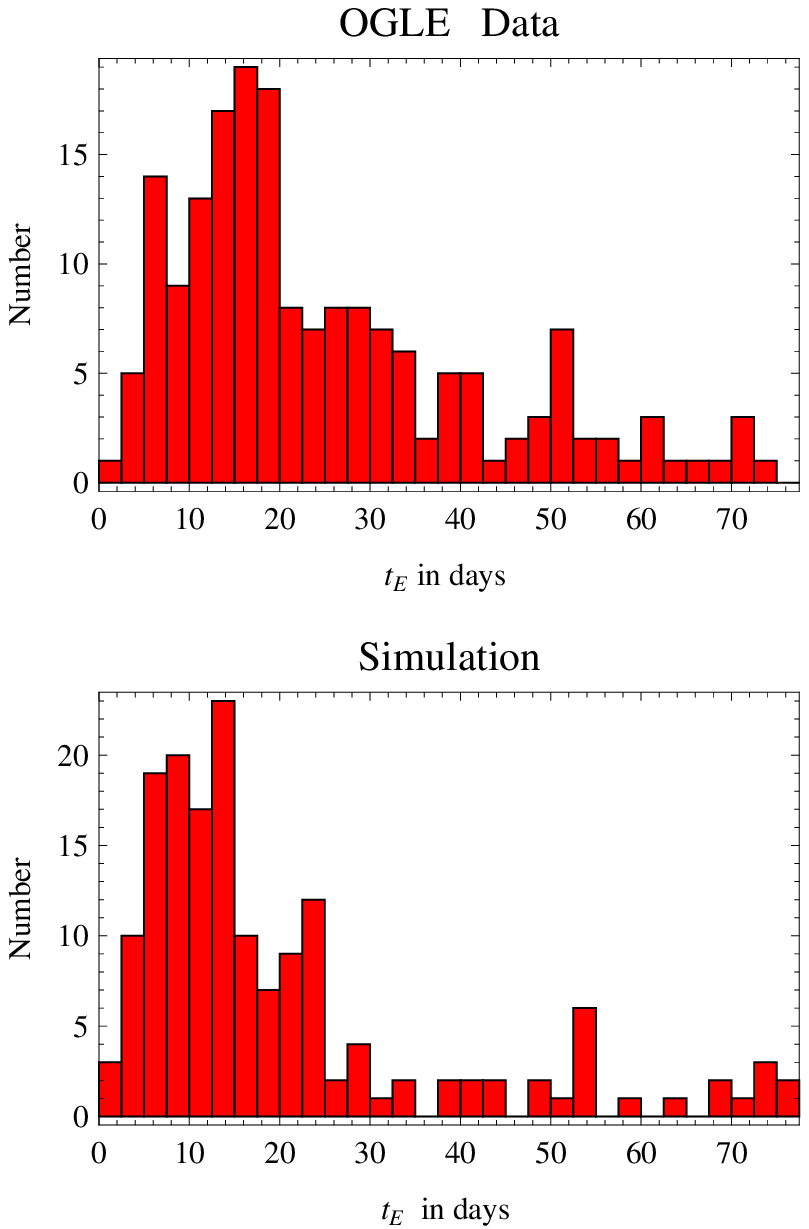, width =   6   in   
}}
\caption{
The data and simulated values  for 
 $t_E$.
Kolmogoroff - Smirnoff
one sample test gives that 
the two sets of data agree with a confidence level greater than 
$97\%$. This is an excellent agreement with our 
hypothesis   that micolensing may be due to both main-sequence stars and 
white dwarfs in the galactic disk. 
}
\end{figure}

\b

There was only very poor agreement if all the source stars
were chosen only from the bar. This is contrary to 
the findings of Alcock et al.$^{8,9}$who had taken all the 
sources to be in the bar. 
Our analysis shows that $90\%$ of the sources could 
come from the bar while the rest must be drawn from a 
disk population. 
Excellent agreement was obtained for all the lens  distances chosen 
equally from the disk and bulk populations. 

\b

The KS test indicate that 
our simulation agrees with the observed distribution with 
significance level greater than 
$97\%$. Our analysis shows  that the possible 
bimodality of $t_E$ could be due to a high percentage 
of white dwarfs($\sim 25\%$) and main-sequence stars.


\begin{thebibliography}{99}

\bibitem{1} Udalski, A., et al., 2000, Acta Astr., 50, 1 

\bibitem{2}Schneider, P. 2006, Extragalactic Astronomy and Cosmology, p. 69

\bibitem{3} Han, C. \&\ Gould, A. ApJ 1995, 447, 53 

\bibitem{4} Scalo, J. M., 1986, Fundamentals of Cosmic Physics, 11, 1

\bibitem{5}Kroupa, P., Tout, C. A., Gilmore, G., 1993, MNRAS, 262, 545

\bibitem{6}DeGennaro, S., arXiv:0709.2190v[astro-ph] 14 Sep 2007

\bibitem{7}Kepler, S. O., et al., 2007, MNRAS, 375, 1315  

\bibitem{8}Alcock, C., et al., 1995, ApJL, 454, L125 

\bibitem{9}Alcock, C., et al., 2000, ApJ, 518, 44 





\end{thebibliography}
\end{document}